\documentclass[letterpaper, 10 pt, journal, twoside]{IEEEtran}

\ifCLASSINFOpdf
\else
\fi

\ifCLASSOPTIONcompsoc
\usepackage{microtype}
 \usepackage[caption=false,font=normalsize,labelfont=sf,textfont=sf]{subfig}
\else
 \usepackage[caption=false,font=footnotesize]{subfig}
\fi
\usepackage{cite}
\usepackage[utf8]{inputenc} 
\usepackage{amsmath,amssymb,amsfonts}
\usepackage{algorithmic}
\usepackage{comment}
\usepackage{graphicx}
\usepackage{textcomp}
\usepackage{xcolor}
\usepackage{booktabs}

\usepackage{scalerel}
\usepackage{tikz}
\usetikzlibrary{svg.path}

\definecolor{orcidlogocol}{HTML}{A6CE39}
\tikzset{
  orcidlogo/.pic={
    \fill[orcidlogocol] svg{M256,128c0,70.7-57.3,128-128,128C57.3,256,0,198.7,0,128C0,57.3,57.3,0,128,0C198.7,0,256,57.3,256,128z};
    \fill[white] svg{M86.3,186.2H70.9V79.1h15.4v48.4V186.2z}
                 svg{M108.9,79.1h41.6c39.6,0,57,28.3,57,53.6c0,27.5-21.5,53.6-56.8,53.6h-41.8V79.1z M124.3,172.4h24.5c34.9,0,42.9-26.5,42.9-39.7c0-21.5-13.7-39.7-43.7-39.7h-23.7V172.4z}
                 svg{M88.7,56.8c0,5.5-4.5,10.1-10.1,10.1c-5.6,0-10.1-4.6-10.1-10.1c0-5.6,4.5-10.1,10.1-10.1C84.2,46.7,88.7,51.3,88.7,56.8z};
  }
}

\newcommand\orcidicon[1]{\href{https://orcid.org/#1}{\mbox{\scalerel*{
\begin{tikzpicture}[yscale=-1,transform shape]
\pic{orcidlogo};
\end{tikzpicture}
}{|}}}}

\usepackage{siunitx}
\DeclareSIUnit{\rad}{rad}
\usepackage{bm}
\def\BibTeX{{\rm B\kern-.05em{\sc i\kern-.025em b}\kern-.08em
    T\kern-.1667em\lower.7ex\hbox{E}\kern-.125emX}}

\usepackage{hyperref}

\hypersetup{
	colorlinks=true,       		
	linkcolor=black,          	
	citecolor=black,        		
	filecolor=black,      		
	urlcolor=black	
}

\begin{document}

\title {
    Impedance Space Method: Time-Independent Parametric Ellipses for Robot Compliant Control 
}

\author {
    Leonardo F. Dos Santos\orcidicon{0000-0003-4836-4543}, Cícero Zanette\orcidicon{0000-0002-0841-7420}, Elisa G. Vergamini\orcidicon{0000-0001-5461-0224}, Lucca Maitan\orcidicon{0009-0002-0763-8675}, and Thiago~Boaventura\orcidicon{0000-0002-9008-9883}
    \thanks{This work was partially supported by São Paulo Research Foundation (FAPESP) under grants 2018/15472-9, 2021/03373-9, 2021/09244-6 and 2023/11407-6, and by Brazilian Federal Agency for Support and Evaluation of Graduate Education (CAPES) under grant 88887.817139/2023-00}
    \thanks{Authors are with the São Carlos School of Engineering, University of São Paulo, Brazil. Email: {\{\tt\footnotesize leonardo.felipe.santos, cicero\_zanette, elisa.vergamini, lucca.maitan, tboaventura\}@usp.br}}
}




\maketitle
\thispagestyle{empty}
\pagestyle{empty}

\begin{abstract}
This paper proposes a novel 3D graphical representation for impedance control, called the \emph{impedance space}, to foster the analysis of the dynamic behavior of robotic compliant controllers. The method overcomes limitations of existing 2D graphical approaches by incorporating mass, stiffness, and damping dynamics, and associates the impedance control parameters with linear transformations to plot a parametric 3D ellipse and its projections in 2D for a mass-spring-damper impedance under sinusoidal reference. Experimental evaluation demonstrates the effectiveness of the proposed representation for analysis of impedance control. The method applies to various compliant control topologies and can be extended to other model-based control approaches.
\end{abstract}


%
\IEEEpeerreviewmaketitle

\section{Introduction} \label{sec:intro}

\IEEEPARstart{R}{obotic} compliant actuation is crucial for safe physical interactions with unknown environments, including humans and assets. Since the 1970s, studies on muscle activation and the central nervous system have introduced mechanical impedance as a key concept, as cited in \cite{grillner1972role}. Drawing from these biological principles, impedance control approaches were formally established a few years later, for example \cite{hogan1985PartI}, providing a viable alternative to traditional position/force control architectures for physical interaction applications. The current state-of-the-art includes a wide range of impedance-based interaction control implementations, spanning machining applications, wearable robots, and transparency control in teleoperation \cite{huo2022adaptive, panzirsch2024enhancing, lloyd2024improved}.

Stiffness, damping, and inertia are the usual dynamic elements composing the desired impedance model. Stiffness and damping are the causally consistent choices based on the hypothesis that the environment is preferably described as an admittance \cite{hogan1985PartI}. The inertia can be the actual system inertia at the interaction port or a desired inertia when applying inertia shaping. Impedance assessment with graphical data usually highlights the spring-damper part of the model. Graphs plot the interaction force against the end-effector's deviation from its equilibrium point \cite{boaventura2015model, song2019tutorial, yoo2019impedance}.

In impedance control, graphical representations such as ellipses and ellipsoids illustrate physical interactions. As described in \cite{focchi2012torque}, the \textit{stiffness ellipse} represents the relationship between deviation and interaction forces in the task space. Each ellipse visually demonstrates the deviation-force relationship between Cartesian directions, such as x and y, when subjected to disturbance forces. The parameters of the diagonal of the Cartesian stiffness matrix define the theoretical ellipse axes, which serve as a reference for the realized task stiffness. This concept is also explored in \cite{ajoudani2015role}, where the limitations of the realized stiffness are analyzed for the robot Jacobian and joint space limits, such as torque saturation. Although this approach has been applied to manipulators and legged robots in these studies, the complexity of three- or six-dimensional impedance formulations and hardware-related constraints prevent the consideration of damping and inertia parameters in this approach.

Graphical representations of stiffness are invaluable for adaptive and variable impedance controllers. In \cite{zhang2023model}, the graphs showing the end-effector position and interaction force tracking are displayed separately over time to evaluate the desired impedance model on the adaptive reference controller. \cite{dalle2024passivity} apply a variable impedance control strategy with an integrated energy tank for passive interaction on industrial collaborative applications with incremental learning from demonstrations. Experimental data shows the end-effector deviation, interaction force, and equivalent stiffness during demonstrations. In contrast, \cite{mazare2022adaptive} present a phase portrait with deviation versus the time derivative of the deviation ($e \times \dot{e}$) to demonstrate the stability of a variable impedance controller applied in a soft robot. The phase plane is the basis for qualitatively characterizing the stability of a dynamic system, identifying stable or unstable points and limit cycles in the state space.

The aforementioned studies use two-dimensional (2D) graphical representations to depict either the deviation-force curve or the deviation and force over time. Additionally, the three-dimensional (3D) graphs illustrate the stiffness ellipsoid, which is the realized stiffness across task space directions. Currently, there is no explicit relationship between impedance model parameters and their graphical representation for oscillatory input in one direction in either task or joint space. Furthermore, to our knowledge, no description encompasses the phase plane, stiffness, and damping behavior simultaneously.

While previous definitions related to ellipses describe the relationship between task-space stiffness or damping in various directions, this work aims to describe impedance dynamics. We clarify the use of the term \textit{ellipse} in impedance control to distinguish our concepts from previous definitions. A parametric representation is useful for descriptive stability and performance analysis. Additionally, it is essential to evaluate the impact of robot dynamics on the controller. This work addresses these issues with the following contributions:

\begin{itemize}
    \item Introducing a novel 3D graphical representation of impedance, termed the \textit{impedance space}, characterized by $e(t) \times \dot{e}(t) \times f_{int}(t)$. This representation extends traditional 2D plots: stiffness ($e(t) \times f_{int}(t)$), damping ($\dot{e}(t) \times f_{int}(t)$), and impedance phase space ($e(t) \times \dot{e}(t)$).
    \item Presenting a 3D elliptic curve in the \textit{impedance space} for mass-spring-damper systems under sinusoidal input. The curve is developed from a 2D unity radius circle and undergoes linear transformations, parameterized by system mass, stiffness, and damping.
    \item Conducting experimental validation of the 3D elliptic curve by constructing the impedance space using real data and employing ellipse fitting for parametric analysis.
\end{itemize}

The focus of this paper is on active compliant control utilizing impedance analogy or similar control topologies specifically for torque-controlled robots. Furthermore, the proposed method has the potential to extend to other areas such as admittance control, physical interaction control, and various model-based methods. Table~\ref{tab:var} aims to aid the reader on following the mathematical process of this paper.

This paper is structured as follows:
Sec. \ref{sec:fundamentals} introduces relevant concepts, the impedance dynamic model, and presents 2D graphs for visualizing stiffness and damping with a sinusoidal input; Sec. \ref{sec:impedance_space} derives the parametric equation of the elliptic path in impedance space and explains the physical rationale behind this representation for evaluating impedance control; Sec. \ref{sec:experimental} details the experimental setup and tests conducted to demonstrate the value of ellipse-based analysis; finally, Sec. \ref{sec:conclusions} summarizes this work and suggests future research directions.

\begin{table}[tb]
    \centering
    \caption{Variables and their meanings}
    \begin{tabular}{cc}
         \toprule
         \textbf{Variable} & \textbf{Meaning} \\
         \midrule
         $t$                            & Time \\
         $e$                            & Position error \\
         $\dot{e}$                      & Velocity error \\
         $f_{int}$                      & Interaction force \\
         $Z$                            & Impedance \\
         $Y$                            & Admittance \\
         $k_d$                          & Desired stiffness \\
         $d_d$                          & Desired damping \\
         $m_d$                          & Desired mass \\
         $a_i$                          & Input amplitude \\
         $\omega_i$                     & Input angular frequency \\
         $\bm g$                        & Tangent unit vector \\
         $\bm n$                        & Normal unit vector \\
         $\bm b$                        & Binormal unit vector \\
         $\bm T$                        & Transformation matrix \\
         $\bm f_1$, $\bm f_2$           & Conjugate diameters \\
         $\varphi$, $\rho$              & 3D rotation angles of $\bm b$ \\
         $\hat{\varphi}$, $\hat{\rho}$  & Projected rotation angles of $\bm b$ \\
         \bottomrule
    \end{tabular}
    \label{tab:var}
\end{table}

\section{Fundamentals} \label{sec:fundamentals}

In a classical perspective, a controller is characterized by its ability to track references and reject disturbances in time or frequency domain. This assessment applies to conventional robot control methods, e.g., force or position control. However, impedance control requires a different perspective. Relying on the physical design paradigm, impedance controllers shapes the interaction dynamics, employing physical reasoning rather than solely adhere to an input-output approach \cite{sharon1991controller, lachner2022shaping}.

That is, the goal is not to track a specific position or force but to control the \emph{dynamic} relationship between a \emph{generalized interaction force} $f_{int}(t)$
and the deviation $e(t)$ from an \textit{equilibrium point} $x_{eq}$ at a designated kinematic location, referred to as the \textit{interaction port}. The equilibrium point can be interpreted simply as the controller reference.

\subsection{Impedance vs Admittance causality}

The impedance and admittance causality are defined in terms of \emph{effort} and \emph{flow} \cite{hogan1987modularity}. Mechanically speaking, an impedance has velocity (flow) as input and force (effort) as output, virtually represented by a massless spring. 
In contrast, an admittance has force as input and velocity as an output, virtually represented by an infinitely rigid mass. In the Laplace domain and for a 1-DoF case, they can be defined as:

\begin{equation}
    Z(s) = \frac{1}{Y(s)} = \frac{F_{int}(s)}{\dot{E}(s)} = \frac{F_{int}(s)}{s{E}(s)} \ ,
    \label{eq:impedanceDefinition} 
\end{equation}

\noindent where $Z(s)$ is the impedance, $F_{int}(s)$ the interaction force, and $E(s) = X_{eq}(s) - X(s)$ the position deviation , being $X_{eq}(s)$ the equilibrium position and $X(s)$ the current position.

\begin{figure}[hb]
    \centering
    \subfloat[][Impedance control block diagram (1-DoF).]{
        \label{fig:impedance_block_diagram_1DoF}%
        \includegraphics[width=0.9\columnwidth]{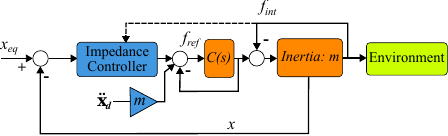}}%
    \qquad \qquad \qquad
    \subfloat[][Impedance control physical equivalence.]{
        \label{fig:impedance_block_diagram_robot}%
        \includegraphics[width=0.65\columnwidth]{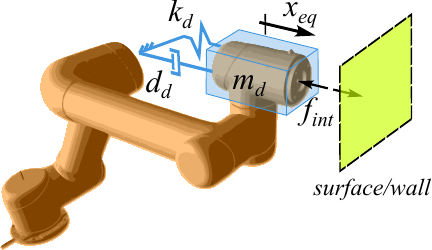}}%
    \caption[]{
    In \subref{fig:impedance_block_diagram_1DoF}, the desired impedance parameters compose the impedance controller blocks in blue. The blocks in orange represent the robot's dynamics that physically interact with the environment (green). The inner torque/force controller $C(s)$ is usually tuned to track with high-fidelity the reference force $f_{ref}$. The interaction force feedback shown by a dashed line highlights its dependency on inertia shaping;
    In \subref{fig:impedance_block_diagram_robot}, a physical equivalence for an impedance controller acting at the end-effector of a manipulator robot.
    }
    \label{fig:impedance_block_diagram}
\end{figure}

As the physical interaction between two systems must be causal, i.e., one system must output what the other system expects as input, an impedance should always be coupled to an admittance, and vice-versa. When the robot interacts with rigid environments, which are preferably admittances, e.g. as shown on Fig. \ref{fig:impedance_block_diagram}, the robot should behave as an impedance at the interaction port \cite{hogan1985PartI, gawthrop2007bond}.

\subsection{System and deviation dynamics}

For a mass-spring-damper designed impedance-controlled system, the impedance model is:

\begin{equation}
    Z(s) = \frac{m_{d}\,s^2+d_{d}\,s+k_{d}}{s} \ ,
    \label{eq:z_s}
\end{equation}

\noindent where $m_d$, $d_d$, and $k_d$ are the desired mass, damping, and stiffness of the system, respectively. Under constrained movement, setting a periodic reference for the designated kinematic location gives the following deviation dynamics:

\begin{equation}
    E(s) = \frac{a_i\,s}{s^2+{\omega_{i}}^2} \ ,
    \label{eq:error_s}
\end{equation}

\noindent where $a_i$ and $\omega_i$ are the amplitude and frequency of the deviation signal, respectively. Applying the Laplace inverse transform the time functions are obtained:
\begin{equation}
    e(t) = \mathbf{L}^{-1}\{E(s)\} = a_i cos(t \omega_i) \ ,
    \label{eq:error_t}
\end{equation}
\begin{multline}
    f_{int}(t) = \mathbf{L}^{-1}\{Z(s) s E(s)\} = \\ = a_i \left(k_d cos(t \omega_i) - d_d \omega_i sin(t \omega_i) - m_d {\omega_i}^2 cos(t \omega_i) \right) \ .
    \label{eq:f_t} 
\end{multline}

\begin{figure}[htb]
    \centering
    \includegraphics[width=1\linewidth]{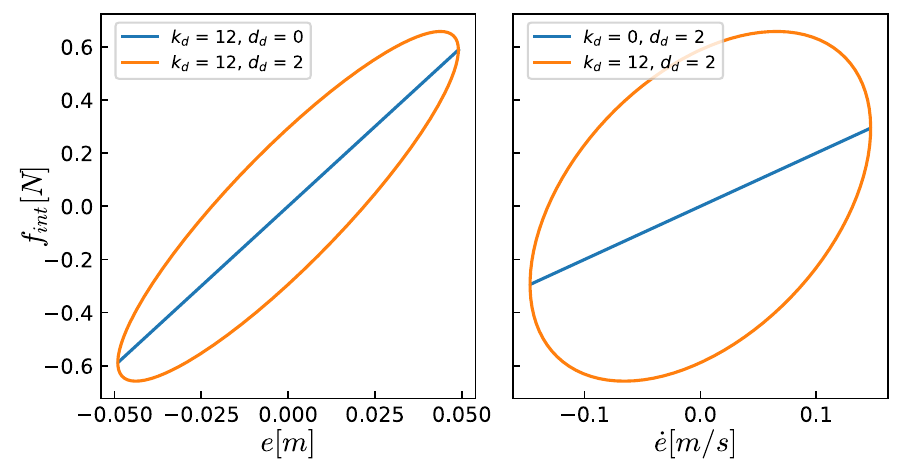}
    \caption{Conventional 2D representation of the impedance stiffness and damping. Values of $k_d$ in \unit{\newton\per\meter} and $d_d$ in \unit{\newton\second\per\meter}. On the left side, on the $e(t) \times f_{int}(t)$ plane, plots for a pure spring, in blue, and a spring-damper, in orange. Note how the inclination of the main axis of the ellipse changes with respect to the line. On the right side, the reciprocal on the $\dot{e}(t) \times f_{int}(t)$ plane.}
    \label{fig:2dGraphs}
\end{figure}

Equations \eqref{eq:error_t} and \eqref{eq:f_t} describe the time-dependent components plot on $e(t) \times f_{int}(t)$ graphs for periodic input. And taking the time derivative of \eqref{eq:error_t} one can see $\dot{e}(t) \times f_{int}(t)$ graph. Figure \ref{fig:2dGraphs} show these graphs. In the first case, the plot is in the stiffness plane, and if only stiffness is set in the impedance controller, the dynamics is linear and given by a line with slope equals to $k_d$; and, if any damping $d_d$ is inserted in the controller, this line turns into an ellipse. The same applies to the second case, where the line slope is equal to $d_d$ and it becomes an ellipse in case of adding a stiffness \cite{charles2012stiffness, RANKO2005169}.

Although these graphs are broadly adopted, no clarification of these 2D ellipses and the impedance model parameters was given before. As it will be clear here, these ellipses at the stiffness and the damping planes are 2D projections of a more complex 3D ellipse.

\section{The Impedance Space} \label{sec:impedance_space}

\begin{figure}[htb]
    \centering
    \includegraphics[width=\linewidth]{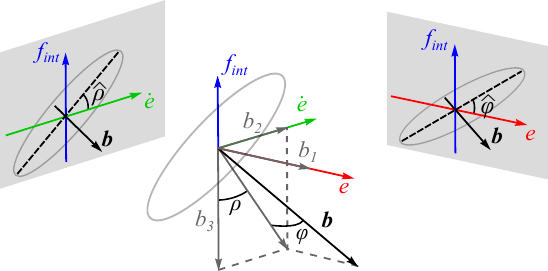}
    \caption{
        The impedance ellipse binormal vector $\bm{b}$, normal to the trajectory plane of a mass-spring-damper impedance under sinusoidal input. The 3D elliptic trajectory is shown in light gray. The components of $\bm{b}$ are $b_1$ for the $e$-axis (red), $b_2$ for the $\dot{e}$-axis (green), and $b_3$ for the $f_{int}$-axis (blue). The left-hand side depicts the projection of the 3D graph on the damping plane, while the right-hand side shows the projection on the stiffness plane. Only in particular cases does the value of $\rho$, and $\varphi$ matches the slope of the 2D ellipse major axis ($\hat{\rho}$, $\hat{\varphi}$) in these planes.
    }
    \label{fig:binormal_vec_angles}
\end{figure}

The \emph{impedance space} is defined using an orthonormal basis. It encompasses three main aspects of impedance dynamics: the stiffness plane ($e \times f_{int}$), the damping plane ($\dot{e} \times f_{int}$), and the phase plane ($e \times \dot{e}$). It is defined as:

\begin{equation}
    \bm{z}(t) = \begin{Bmatrix} e(t) \\ \dot{e}(t) \\ f_{int}(t) \end{Bmatrix} = e(t)\mathbf{i} + \dot{e}(t)\mathbf{j} + f_{int}(t)\mathbf{k} \ ,
    \label{eq:impedanceSpaceBasis}
\end{equation}

\noindent where $e(t)$ and $f_{int}(t)$ are position deviation and interaction force, respectively, and $\bm{i}$, $\bm{j}$, and $\bm{k}$ are the basis vectors.  
In the context of second-order autonomous (implicit time-dependent) systems, represented by $\dot{\bm{x}} = f(\bm{x})$, the phase plane is the foundation for qualitative characterization of system stability through the phase portrait. Stable and unstable points, limit cycles, and multiple isolated equilibrium points are identified through the phase portrait \cite{khalil2002nonlinear, verhulst2006nonlinear}.

For a 2\textsuperscript{nd} order dynamic system, as \eqref{eq:z_s}, under the periodic input \eqref{eq:error_s}, the time-dependent system trajectory in the impedance space lies in a plane. Consider the Frenet-Serret \cite{stewartchap13} triad: a unit vector $\bm{g}(t) = \bm{\dot{z}}(t)/\lVert \bm{\dot{z}}(t)\rVert \in \mathbb{R}^{3}$ tangent to the curve, a normal unit vector $\bm{n}(t) = \bm{\dot{t}}(t)/\lVert \bm{\dot{t}}(t)\rVert \in \mathbb{R}^{3}$ pointing to the center of the curve at $t > 0$ , the binormal unit vector $\bm{b}(t)$, orthogonal to the trajectory, is:

\begin{align}
    \bm{b}(t) = \bm{g}(t) \times \bm{n}(t) = 
    \begin{Bmatrix}\dot{e}(t) \\ \ddot{e}(t) \\ \dot{f}_{int}(t) \end{Bmatrix} 
    \times \frac{d}{dt} \begin{Bmatrix}\dot{e}(t) \\ \ddot{e}(t) \\ \dot{f}_{int}(t) \end{Bmatrix} \ ,
\end{align}

Using \eqref{eq:error_t} and \eqref{eq:f_t}, vector $\bm{b}$ simplifies to:

\begin{equation}
    \bm{b} = \begin{Bmatrix}
        b_1 \\ b_2 \\ b_3
    \end{Bmatrix} \
    = \frac{1}{\sqrt{\left(k_d - m_d\omega_i^2\right)^2 + d_d^2 + 1}}\begin{Bmatrix}
        k_d - m_d\omega_i^2 \\ d_d \\ -1
    \end{Bmatrix} \ .
\end{equation}

Through the orientation of $\bm{b}$ in the impedance space, Fig. \ref{fig:binormal_vec_angles}, the sequential rotation angles $\varphi$ and $\rho$ are defined as follows:
\begin{equation}
    \tan(\varphi) = -\frac{b_1}{\sqrt{b_2^2 + b_3^2}} = \frac{m_d\omega_i^2 - k_d}{\sqrt{d_d^2+1}} \ , \label{eq:phi}
\end{equation} 
\begin{equation}
    \tan(\rho) = -\frac{b_2}{b_3} = d_d \label{eq:rho} \ ,
\end{equation}
\noindent where the negative sign in (\ref{eq:rho}) and (\ref{eq:phi}) is due to the right-handed coordinate systems convention.
As sequential rotation angles, $\varphi$ and $\rho$ have equivalent rotation matrices $\bm{T_e}(\rho), \bm{T_{\dot{e}}}(\varphi) \in \mathbb{R}^{3 \times 3}$. $\bm{T_e}(\rho)$ is a rotation around the $e$-axis. $\bm{T_{\dot{e}}}(\varphi)$ is a rotation around the $\dot{e}$-axis in the base rotated by $\bm{T_e}(\rho)$. Supposing that $z(t)$ can be parametrized by $\theta = \omega_i t \hspace{1pt}, \in [0,2\pi]$, and that in the impedance space the system is described by the following equation:
\begin{equation}
    \bm{z}(\theta) = 
    \bm{T_e}(\rho) \bm{T_{\dot{e}}}(\varphi) \bm{T_{f_{int}}} 
    \begin{Bmatrix}\cos(\theta) \\ \sin(\theta) \\ 0\end{Bmatrix} \ ,
    \label{eq:parametric_transforms}
\end{equation}
\noindent where $\bm{T_{f_{int}}}$ is defined as:
\begin{align}
    \bm{T_{f_{int}}} &=
    \begin{bmatrix}
        \begin{matrix} \bm{R} \end{matrix} & \begin{matrix} 0 \\ 0\end{matrix} \\
        \begin{matrix} 0 & 0  \end{matrix} & 1
    \end{bmatrix} \ ,
\end{align}
\noindent being $\bm{R} = \begin{bmatrix} \bm{f_1} & \bm{f_2} \end{bmatrix} \in \mathbb{R}^{2 \times 2}$ a matrix of the ellipse conjugate diameter $\bm{f_1},\bm{f_2} \in \mathbb{R}^{2}$. Solving \eqref{eq:parametric_transforms} for $\bm{T_{f_{int}}}$, the hypothesis are valid and defines:
\begin{equation}
    \bm{R} =
    \begin{bmatrix}
        a_i \sqrt{\frac{(k_d - m_d \omega_i^2)^2}{d_d^2 + 1} + 1} & 0 \\
        \frac{a_i d_d (k_d - m_d \omega_i^2)}{\sqrt{d_d^2 + 1}} & -a_i \omega_i \sqrt{d_d^2 + 1}
    \end{bmatrix} \ .
    \label{eq:Tz2x2}
\end{equation}

Then, it can be stated that the impedance model \eqref{eq:z_s} under the input \eqref{eq:error_s} describes a parametric 3D elliptic curve in the impedance space. This curve is described by a sequence of affine transformations upon the 2D parametric unit circle $[x\,,\,y]  = [\cos(\theta)\,,\,\sin(\theta)]$, with $\theta \in \left[0, 2\pi\right]$ in the form of the Equation \ref{eq:parametric_transforms}.

While the first two matrices are rotations, $\bm{T_{f_{int}}}$ scales and shears the parametric unit circle in the phase plane. Since the $f_{int}$-axis is normal to this plane, this transformation is named after it. With these matrices, the 3D elliptic curve can be derived without time dependence, being parameterized only by $\theta$, the model constants and the input parameters $\omega_i$ and $a_i$.

\begin{figure*}[htb]
    \centering
    \subfloat[][]{
        \label{fig:ellipses_colored_D}
        \includegraphics[width=0.235\textwidth]{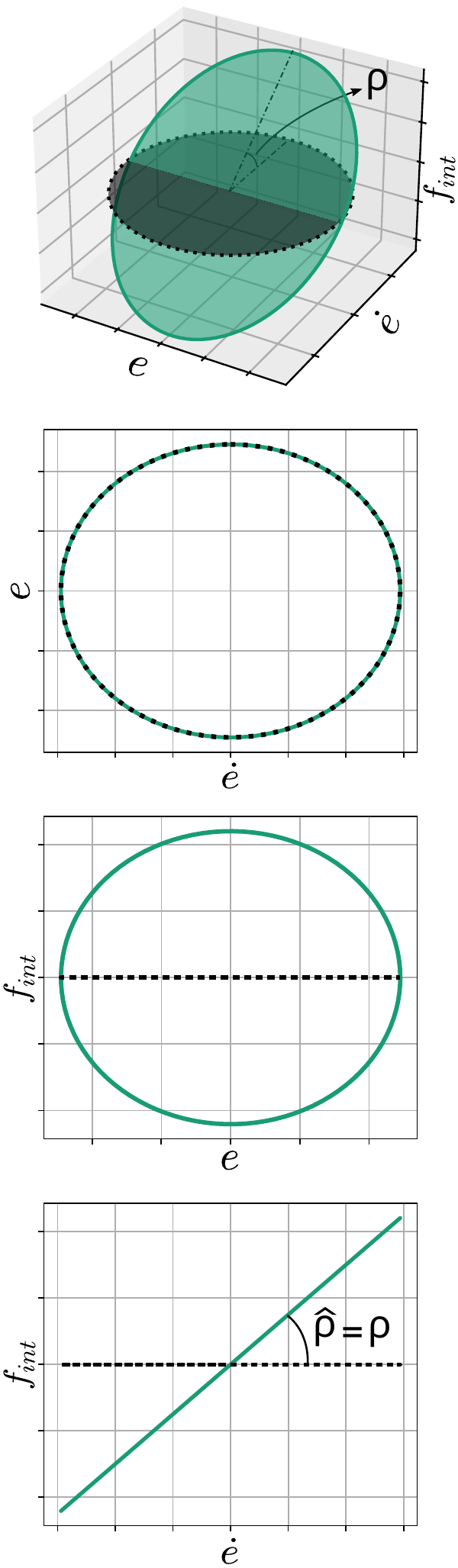}
    }
    \subfloat[][]{
        \includegraphics[width=0.235\textwidth]{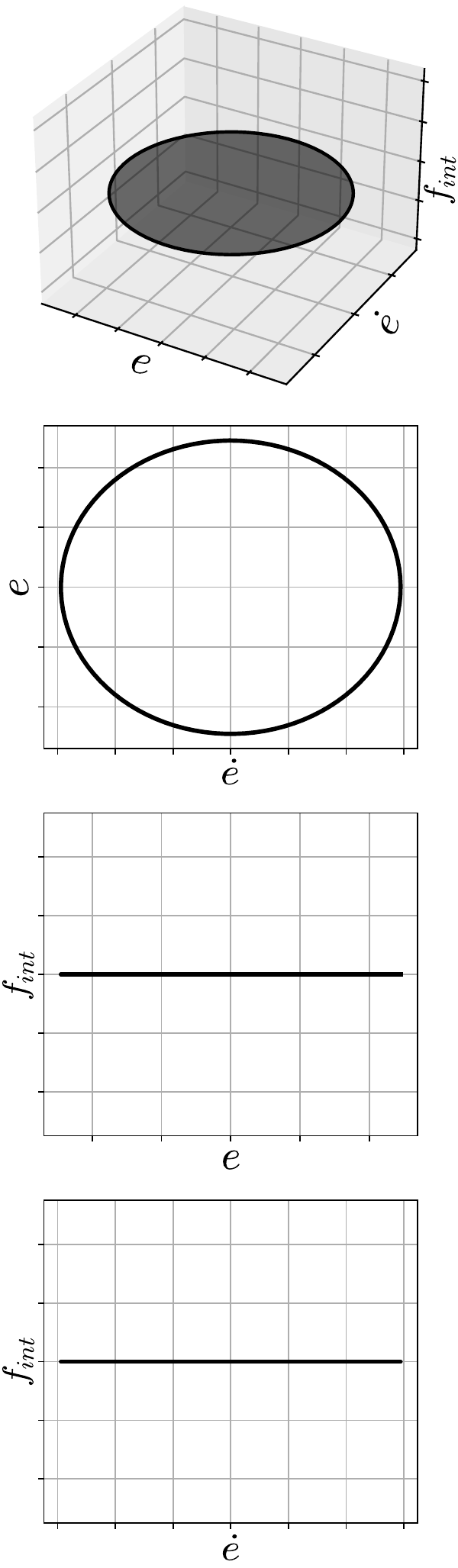}
        \label{fig:ellipses_colored_0}
    }
    \subfloat[][]{
        \includegraphics[width=0.235\textwidth]{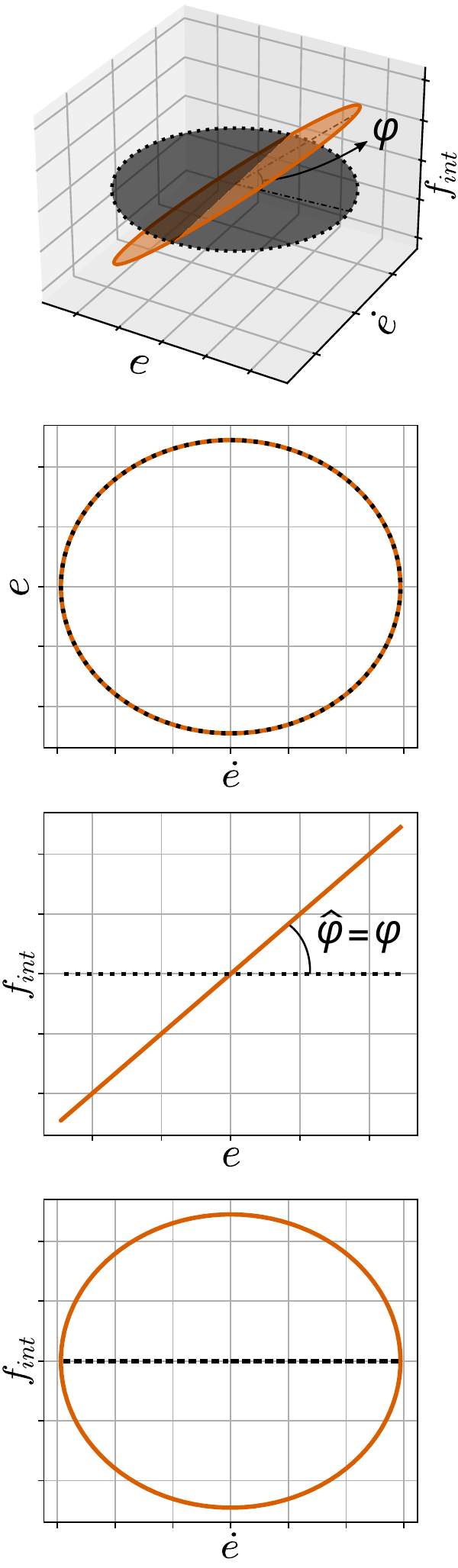}
        \label{fig:ellipses_colored_K}
    }
    \subfloat[][]{
        \includegraphics[width=0.235\textwidth]{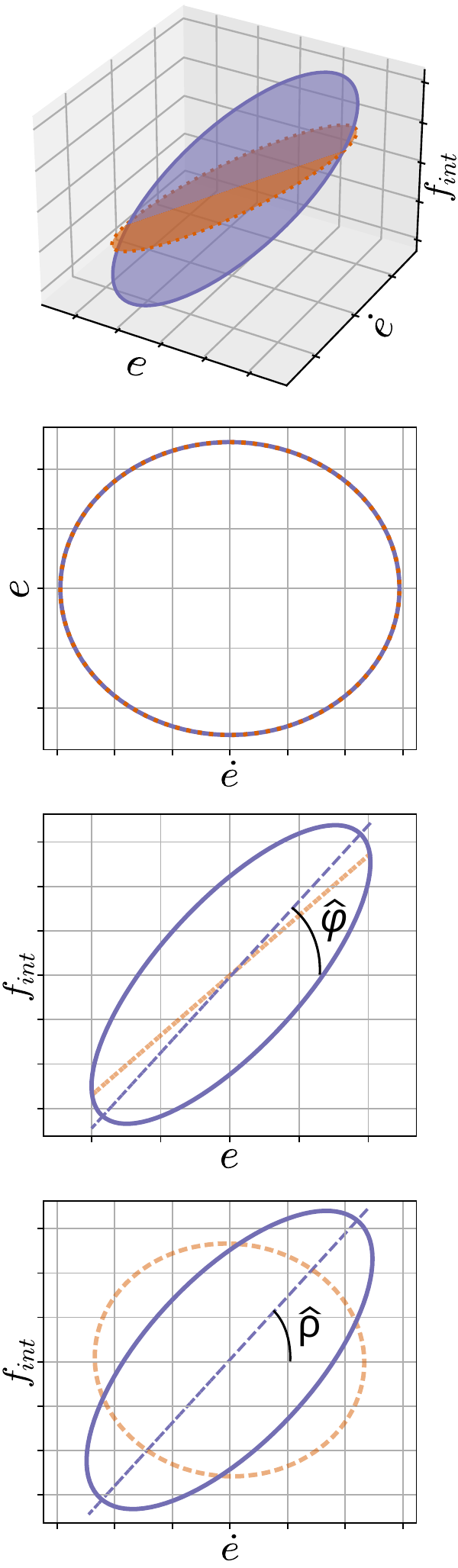}
        \label{fig:ellipses_colored_KD}
    }
    \caption[]{Visualization of the 3D theoretical impedance plots and its projections. Each column represents a specific combination of impedance parameters $k_d$ and $d_d$ ($m_d=0$ for all of them), while the rows show different projection of the 3D impedance plot for each case. The first row, in particular, shows the 3D \emph{impedance space}.
    In \subref{fig:ellipses_colored_0}, where $k_d=0$ and $d_d=0$, only the original unit circle is plotted, from which the other columns derive.
    In \subref{fig:ellipses_colored_D}, $d_d \neq 0$ is added, and the original circle, in black, is elongated and then rotated by $\rho$ around the $e$ axis. An ellipse appears in the 3D plot, with a projection on the plane $\dot{e} \times f_{int}$ being a straight line with the slope given by $tan(\hat{\rho})=tan(\rho)=d_d$.
    In \subref{fig:ellipses_colored_K}, $k_d \neq 0$ is added to the original circle in \subref{fig:ellipses_colored_0}, which is elongated and then rotated by $\varphi$ around the $\dot{e}$ axis in this case. The 2D ellipse projection on $e \times f_{int}$ is also a straight line, with a slope given by $tan(\hat{\varphi})=tan(\varphi)=k_d$.
    Last but not least, in \subref{fig:ellipses_colored_KD}, the original circle is transformed into a 3D ellipse with $k_d \neq 0$ and $d_d \neq 0$. In this case, the 2D projections in both $e \times f_{int}$ and $\dot{e} \times f_{int}$ planes are also ellipses. It is important to highlight that, for these 2D projected ellipses, $tan(\hat{\varphi}) \neq k_d$ and $tan(\hat{\rho}) \neq d_d$. In particular, in the $e \times f_{int}$ plane it is possible to see how the inclination changes with respect to the pure stiffness case of \subref{fig:ellipses_colored_K}.}
    \label{fig:ellipses_colored}
\end{figure*}

Fig.~\ref{fig:ellipses_colored} shows the isometric and perspective views of the impedance ellipse and its dependence on the impedance parameters $k_d$ and $d_d$. The projection of the rotation angles $\rho$ and $\varphi$ into the $\dot{e} \times f_{int}$ and $e \times f_{int}$ planes is represented by $\hat{\rho}$ and $\hat{\varphi}$, respectively. The analytical expressions of these angle projections are very complex, highly nonlinear and dependent on all the desired impedance parameters\footnote{Full expressions at \url{https://github.com/leggedrobotics-usp/impedance\_control\_benchmark}}. Only in particular cases, the value of these projections simplify to $\hat{\rho}=\rho$, as depicted in Fig. \ref{fig:ellipses_colored_K}, and $\hat{\varphi}=\varphi$, in Fig. \ref{fig:ellipses_colored_D}.

\begin{figure}[tb]
        \centering
        \includegraphics[width=1\columnwidth]{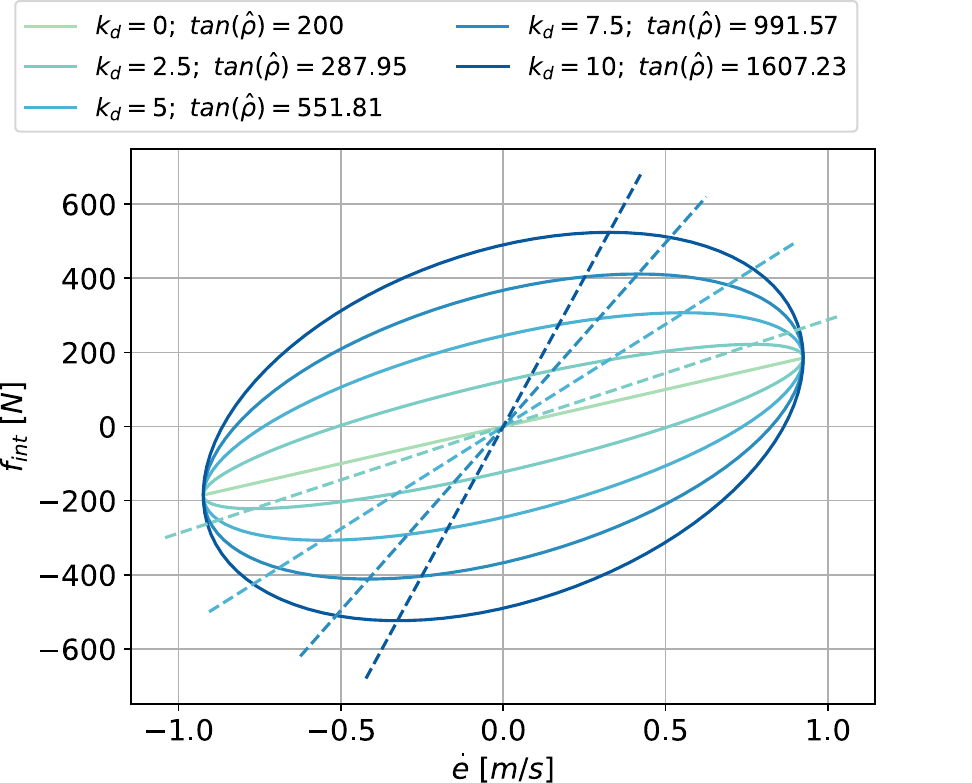}
        \caption{Plot of the 2D projected ellipse in the $\dot{e} \times f_{int}$ plane, with a fixed $d_d=200$~\unit{\newton\second\per\meter}. Values of $k_d$ in \unit{\newton\per\meter}.
        When a stiffness $k_d$ is added to the impedance controller, the slope of the ellipse major axis $tan(\hat{\rho})$ may significantly differ from $d_d$.}
        \label{fig:KinfluenceRho}
\end{figure}

\begin{figure}[htb]
        \centering
        \includegraphics[width=1\columnwidth]{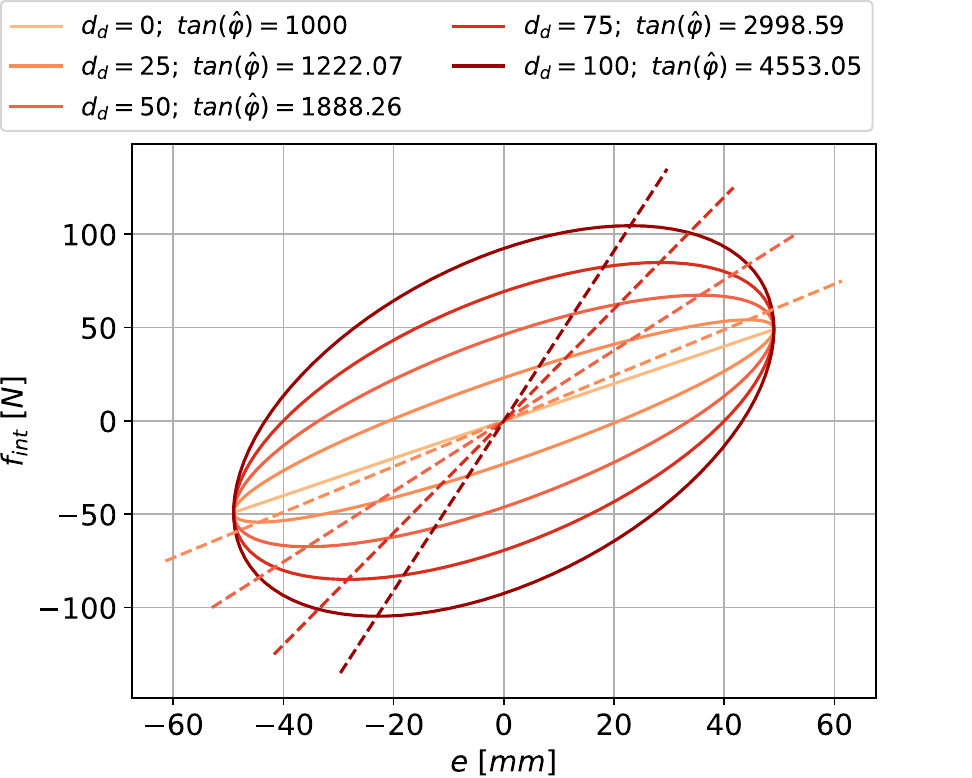}
        \caption{Plot of the 2D projected ellipse in the $e \times f_{int}$ plane, with a fixed $k_d=1000$~\unit{\newton\per\meter}. Values of $d_d$ in \unit{\newton\second\per\meter}. 
        When a damping $d_d$ is added to the impedance controller, the slope of the ellipse major axis $tan(\hat{\varphi})$ may significantly differ from $k_d$.} 
        \label{fig:DinfluencePhi}
\end{figure}

\begin{figure*}[tb]
        \centering
        \includegraphics[width=2\columnwidth]{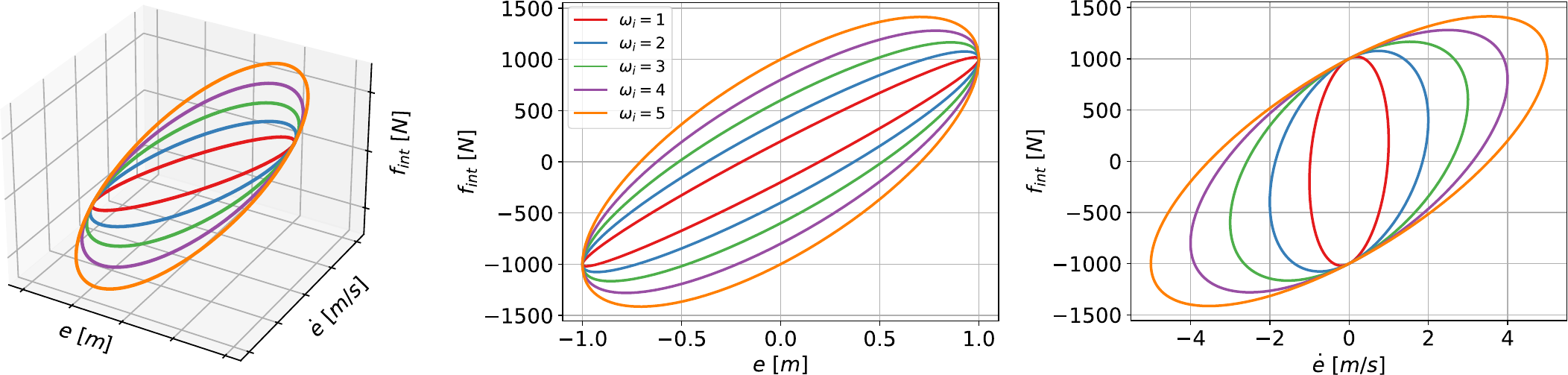}
        \caption{Influence of the input frequency $\omega_i$, in \unit{\rad\per\second}, on the impedance ellipses for the parameters $m_d=0$~\unit{\kilogram}, $k_d=1000$~\unit{\newton\per\meter} and $d_d=200$~\unit{\newton\second\per\meter}.}
        \label{fig:wi}
\end{figure*}

This new approach elucidate the interpretation of the 2D ellipses, and the apparent slopes in the 2D projections. Figures \ref{fig:KinfluenceRho} and \ref{fig:DinfluencePhi} show how the slope changes according the sitffness and the damping, respectively. Moreover, the influence of the input frequency $\omega_i$, is shown in Fig.~\ref{fig:wi}. Even for the case $m_d=0$~\unit{\kilogram}, where $w_i$ affects only the $\bm{T_{f_{int}}}(2,2)$ term, the impedance ellipse changes significantly the shape.

\begin{figure}[htb]
    \centering
    \includegraphics[width=1\linewidth]{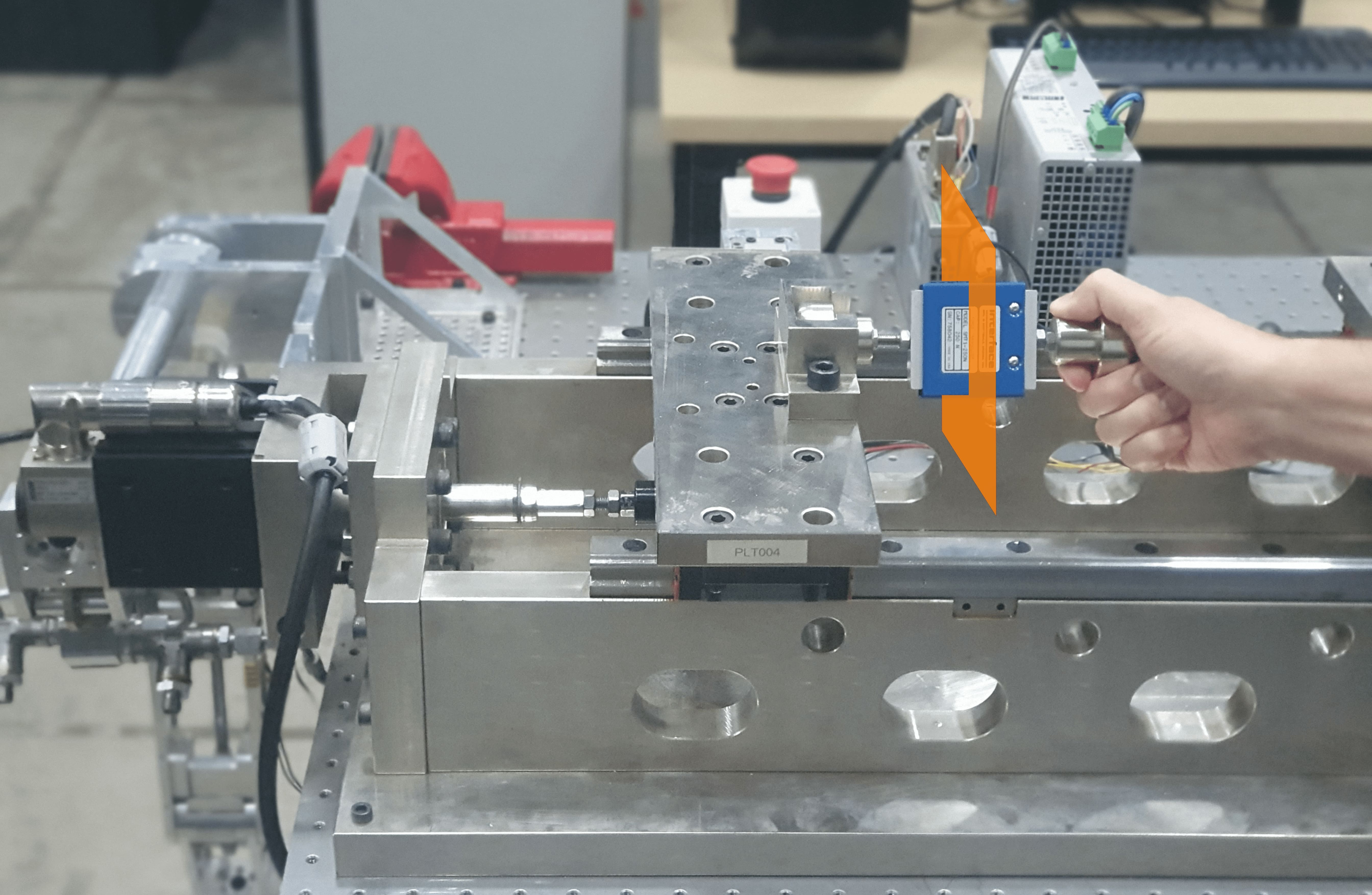}
    \caption{Experimental setup. The interaction port is highlighted in orange. The interaction force was provided by a human hand, and a linear encoder measured the position. From left to right: linear motor, platform, load cell. The encoder is behind the platform, measuring its position.}
    \label{fig:ic2d}
\end{figure}

\section{Experimental impedance ellipses} \label{sec:experimental}

\begin{figure}[t]
        \centering
        \includegraphics[width=0.9\columnwidth]{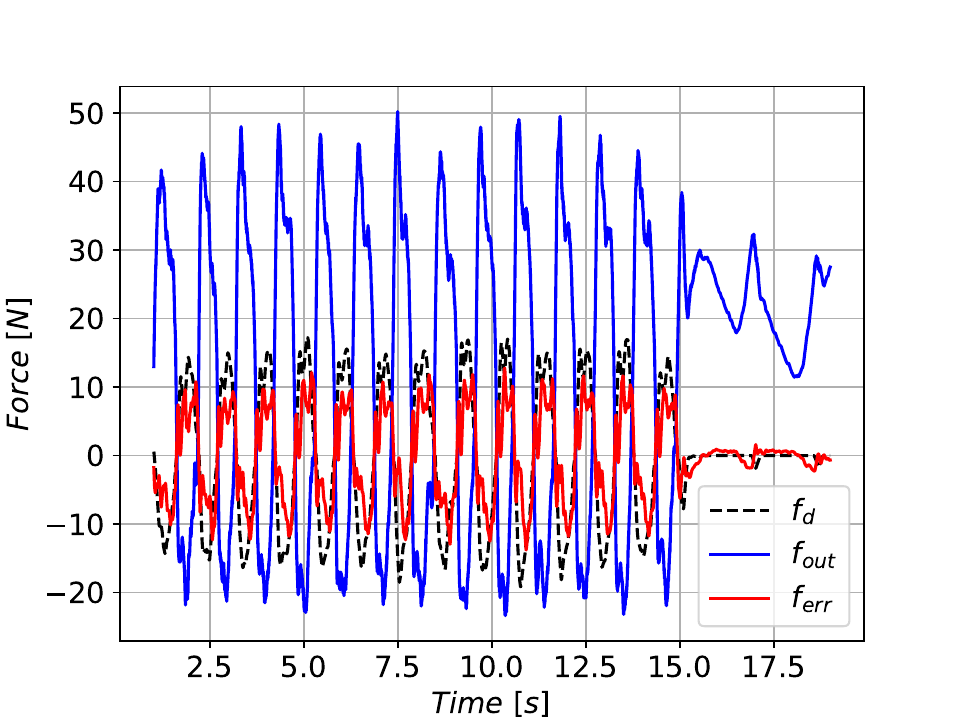}
        \caption{Experimental data showing the reference tracking of the inner force controller for the experiment with $k_d=0$ \unit{\newton\per\meter} and $d_d=200$ \unit{\newton\second\per\meter}. The plot shows the desired interaction force ($f_d$, black), the force error ($f_{err} = f_{int} - f_d$, red), and the controller output ($f_{out}$, blue). The controller output is expressed in Newtons and corresponds to the force command; this value is later converted into motor current using the actuator's force-to-current constant. This force tracking performance influences the rendered impedance.}
        \label{fig:Forcetracking}
\end{figure}

The experimental data was collected using a specially designed test bench named IC2D, shown in Fig.~\ref{fig:ic2d}. The linear motor used in the experiment was the LinMot PS01-37x120F-HP-C (stator) and PL01-20x240 / 180HP (slider) models. The interaction force was measured using an Interface SMT1-250N load cell and and for the position we used the LM10IC001AB10F00 incremental encoder from RLS. A detailed description of the experimental bench is provided in \cite{Gamper_Vergamini_2023, IC2D2023}.  In both 2D and 3D, the impedance ellipse may also be plotted using experimental data to form the impedance state vector $z(t)$. The inner force controller, implemented in cascade with the impedance controller, was a PID with friction compensation through disturbance observer-based control (DOB) and load velocity compensation  \cite{DOB_Calanca,Vel_compe_thiago}. Its tracking performance is shown in Fig. \ref{fig:Forcetracking}.

Nine experiments were conducted to demonstrate the parametric ellipses according to the parameters $k_d$ and $d_d$. The interaction port was manually excited with a sinusoidal force. Despite the repeatability of the test bench, the accuracy of this parametric analysis was found to be sensitive to system force tracking fidelity and joint friction. A statistical treatment of the data is beyond the scope of this work. The experimental results support the present method qualitatively.

\begin{figure*}[tb]
    \centering
    \subfloat[][]{
        \includegraphics[width =0.29999952\textwidth]{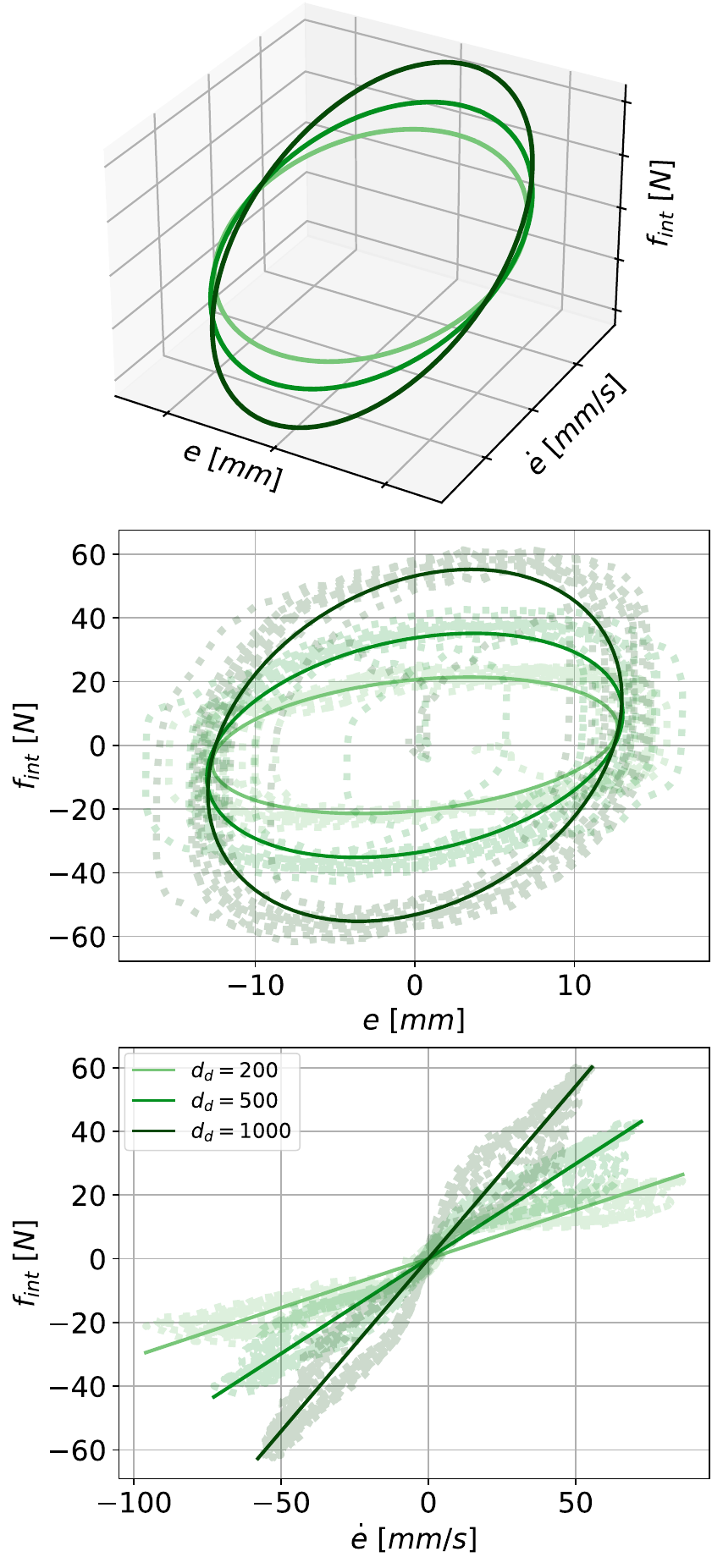}
        \label{fig:D_variation}
    }
    \subfloat[][]{
        \includegraphics[width=0.31\textwidth]{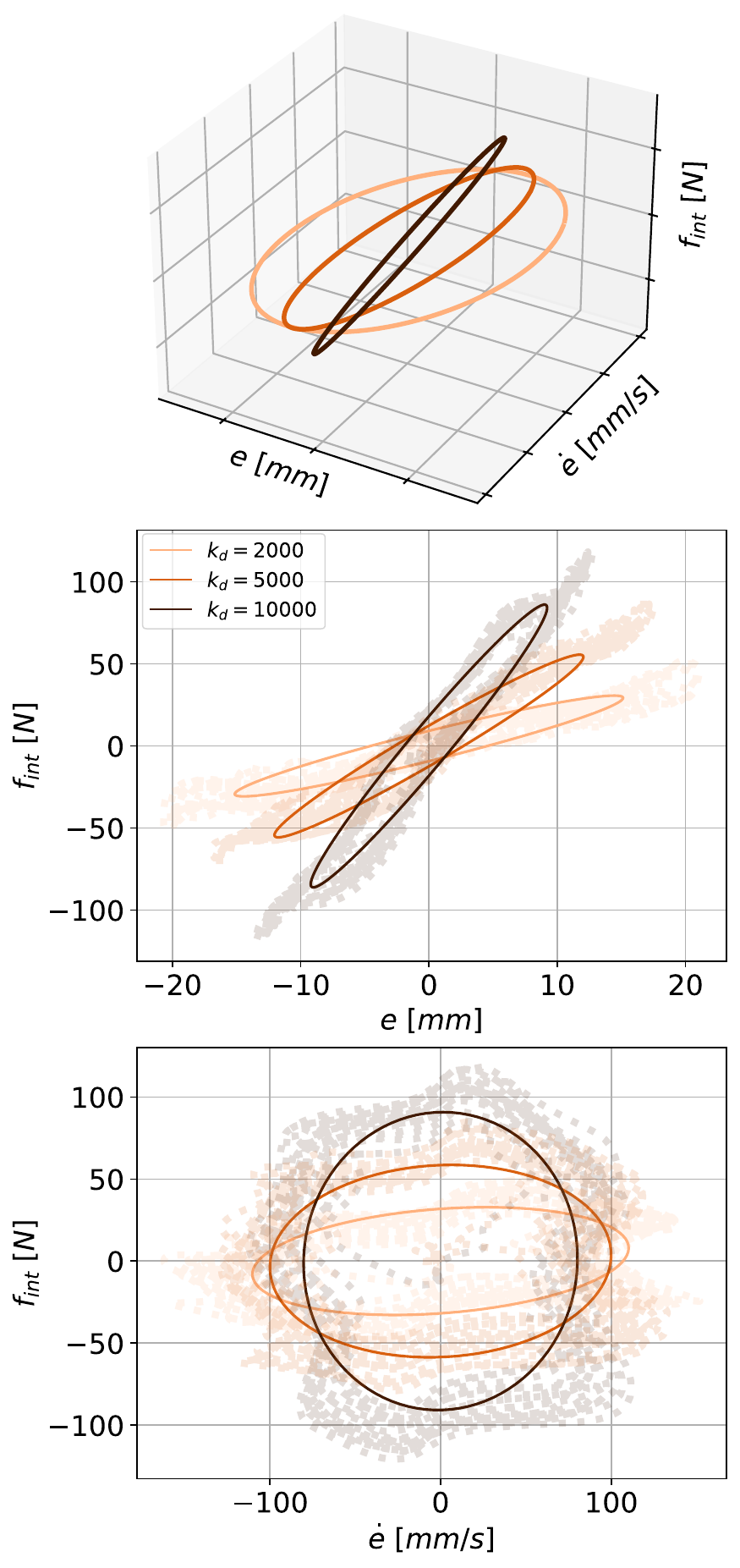}
        \label{fig:K_variation}
    }
   \subfloat[][]{
        \includegraphics[width=0.31\textwidth]{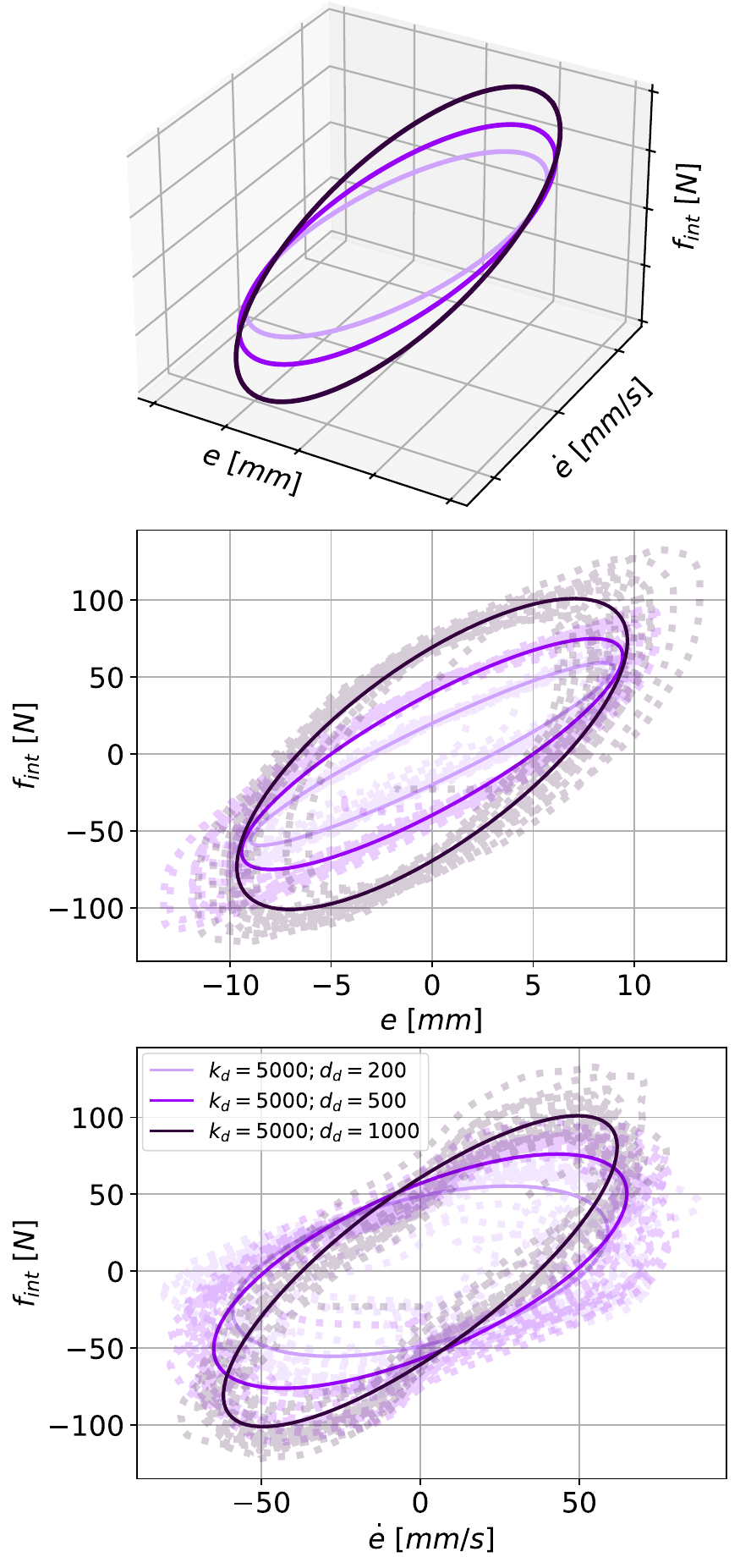}
        \label{fig:Kfix_Dvariation}
   }
        \caption[]{Experimental results, all with $m_d = 0$ \unit{\kilogram}.
        \subref{fig:D_variation} for $k_d = 0$ \unit{\newton\per\meter} and varying the value of $d_d$ one can see that in the $\dot{e} \times f_{int}$ plane the straight line slope increases with $d_d$ and in the $e \times f_{int}$ plane the ellipse widens due to the projection as expected;
        \subref{fig:K_variation} for $d_d = 0$ \unit{\newton\second\per\meter} and varying the value of $k_d$, one can see that in the $e \times f_{int}$ plane, the slope of the ellipse major axis increases with $k_d$ and in the $\dot{e} \times f_{int}$ plane the ellipse widens due to the projection as expected. The widening of the ellipse in the $e \times f_{int}$ plane is due to unwanted damping, mostly caused by the test bench itself. As can be seen, the openings are practically the same, as the damping of the bench does not change;
        \subref{fig:Kfix_Dvariation} for both parameters $k_d$ and $d_d$ varying, the effects combine on both planes.}
    \label{fig:experimental_variation}
\end{figure*}

Figure~\ref{fig:experimental_variation} shows the fitted ellipses for each parameter set, utilizing the stable direct least squares fitting algorithm from Halíř and Flusser \cite{halir1998numerically}. Although the ellipse fit helps identify the rendered impedance, stiffness, and damping values, the performance of the inner-loop force tracking of the linear motor and the friction of the test bench affected the experimental data. Nevertheless, the ellipse fit captured the relevant graphical aspects in impedance space.

Using the impedance space allows new metrics to assess the rendered impedance against the desired one 3D ellipse. Examples include: 1) ``interaction force fidelity'' measures differences in force magnitudes for each specific point between ellipses; 2) ``impedance 3D orientation error' compares the binormal vectors of ellipse planes, incorporating mass, damping, and stiffness; 3) ``stiffness error'' and ``damping error'' are specific metrics easily derived by comparing  \autoref{eq:phi} (for $m_d=0$) and \autoref{eq:rho}, respectively, which are angles defined in the 3D space, rather than relying on their 2D projections.

Moreover, in \cite{ott2008cartesian} was present a stiffness plot, where the data presented a so called \textit{hysteresis}-like shape which is attributed to the damping only. However, as it is clear now, this shape correlation with damping is described by an ellipse projection, not a hysteresis phenomenon which is related to friction forces.

\section{Conclusions \& Outlook} \label{sec:conclusions}
This paper introduced and discussed a novel representation of mechanical impedance ellipses in a 3D impedance space $e(t) \times \dot{e}(t) \times f_{int}(t)$. A 3D parametric elliptical curve describes the impedance dynamics using linear transformations, which are functions not only of the impedance controller parameters $m_d$, $k_d$, and $d_d$, but also of the input amplitude $a_i$ and frequency $\omega_i$. This modeling and representation approach fosters parameter-oriented analysis of experimental time series using the designed dynamic behavior as a reference. The impedance rendering performance obtained with experimental data in this paper was limited by the inner force controller's tracking capabilities and the amplitude and frequency input fluctuations within each cycle due to the manual excitation. These factors also affect systems with a higher number of degrees of freedom. For n-DoF systems, one challenge is associating joint-level force tracking with the robot Jacobian and the impedance space trajectory for a given direction in task space.
As a final remark, this paper highlights that the phase space $e(t) \times \dot{e}(t)$, included in the presented impedance space, supports further nonlinear and energy-based analysis. Future developments include calculating the mentioned metrics, extending the analysis including the impedance state $\ddot{e}(t)$ and the influence of inertia shaping.

 \section*{Acknowledgment}

The authors would like to thank the Legged Robotics Group of the Robotics Laboratory of the São Carlos School of Engineering, University of São Paulo.

\bibliographystyle{bibtex/bib/IEEEtran}
\bibliography{bibtex/bib/references}

\newpage
\begin{IEEEbiography}[{\includegraphics[width=1in,height=1.25in,clip,keepaspectratio]{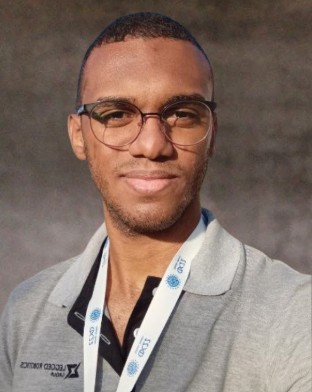}}]{Leonardo F. dos Santos} holds a degree in Mechatronics Engineering from the São Carlos School of Engineering at the University of São Paulo (2021). His interests lie in robotics, systems modeling and control, both in academia and industry. Leonardo is currently a member of the Legged Robotics Group and is pursuing is PhD within the FAPESP Young Researcher Project entitled 'Impedance control of hydraulic actuators for robots with legs and arms'. He is also a member of SENAI CIMATEC's robotics division, where he contributes to initiatives focusing on robotic manipulation for underwater inspection, maintenance, and repair.
\end{IEEEbiography}

\vspace{-33pt}

\begin{IEEEbiography}[{\includegraphics[width=1in,height=1.25in,clip,keepaspectratio]{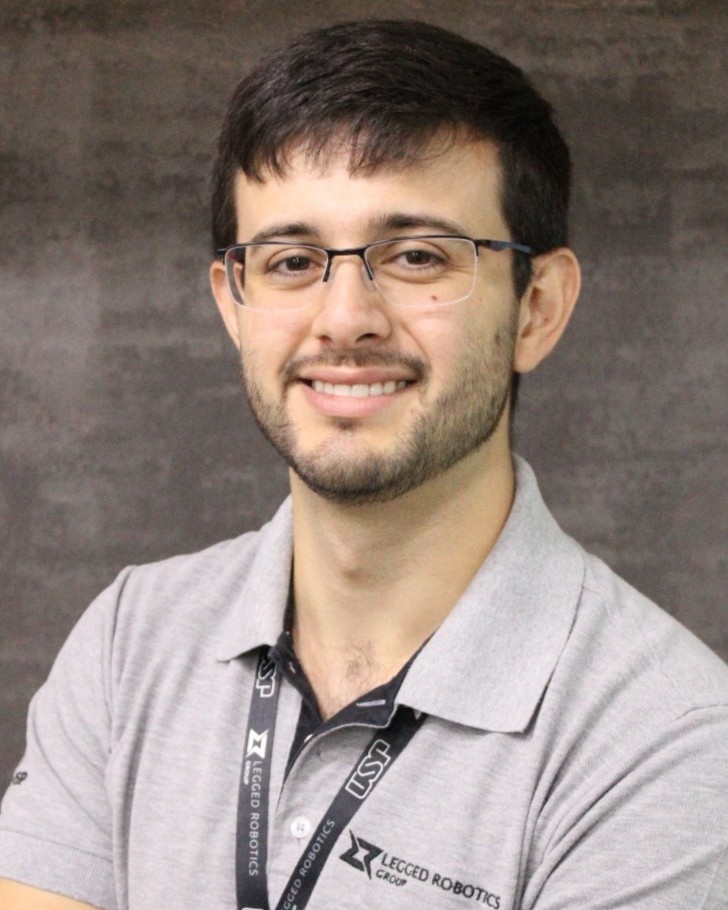}}]{Cícero Zanette}
received his B.Sc degree in Mechanical Engineering from the Federal University of São Carlos, Brazil, in 2021. During his undergraduate studies, he worked with computational simulations in the area of induced cracks in high-hardness ceramics during laser-assisted machining processes utilizing FEM. He received his M.Sc degree in Dynamics and Mechatronics from the São Carlos School of Engineering of the University of São Paulo, Brazil, where he worked with the Legged Robotics Group in the development of metrics for comparison of hybrid impedance and admittance controllers. He is currently pursuing his D.Sc. degree in the same research group.
\end{IEEEbiography}

\vspace{-33pt}

\begin{IEEEbiography}[{\includegraphics[width=1in,height=1.25in,clip,keepaspectratio]{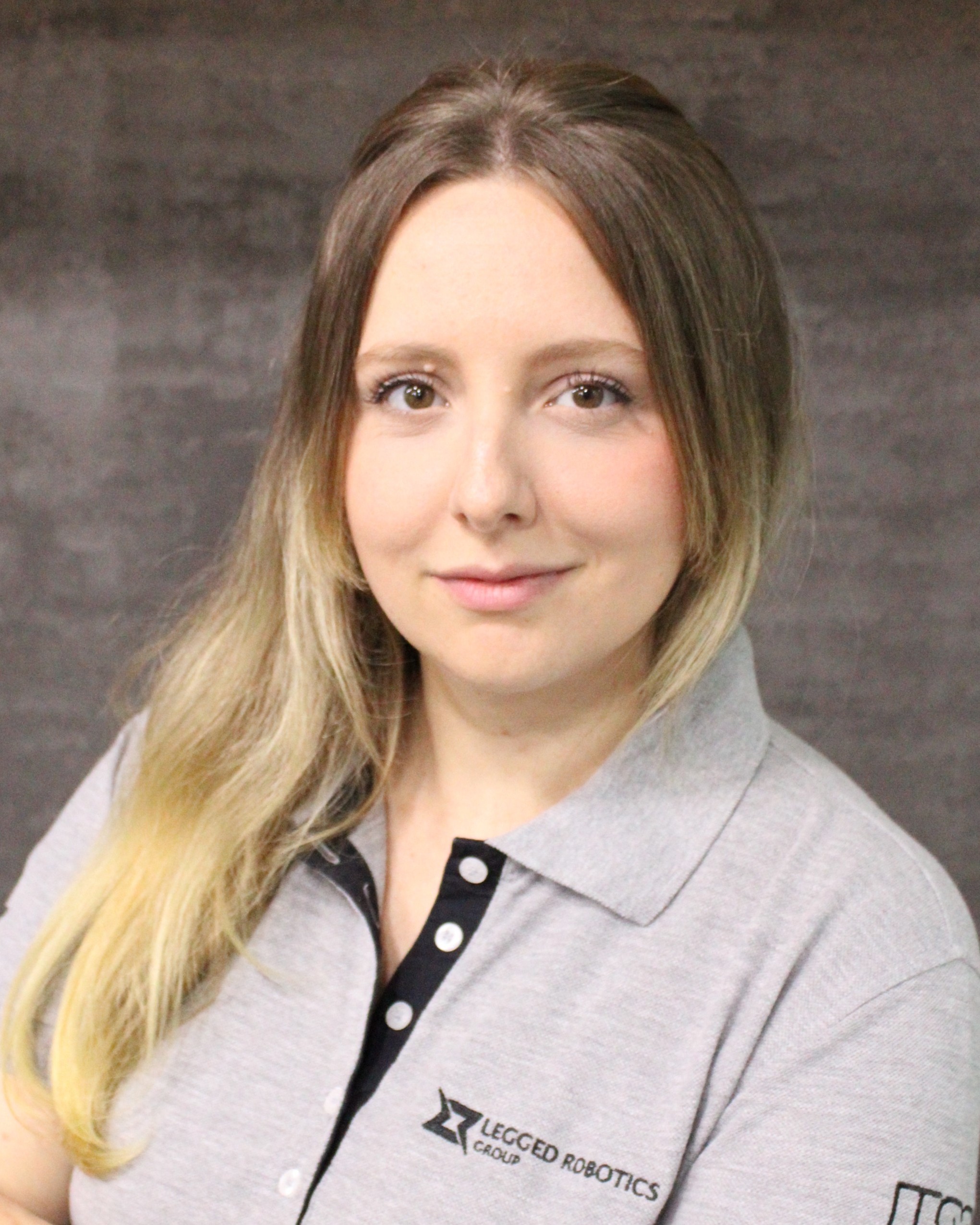}}]{Elisa G. Vergamini}
graduated in Mechanical Engineering from the Federal University of São Carlos, Brazil, in 2020. In 2018, Elisa studied for a year at the École Nationale Supérieure de Techniques Avancées, in the undergraduate course in Mechanical Engineering with an emphasis on Intelligent Systems. In 2022, she carried out a research internship abroad for six months (BEPE - FAPESP) at the Altair Robotics Lab at the University of Verona. In 2024, Elisa completed her master's degree at the Robotics Laboratory of the Department of Mechanical Engineering of the São Carlos School of Engineering of the University of São Paulo, at the Legged Robots Group. She is currently pursuing her PhD in the same Research Group.
\end{IEEEbiography}

\vspace{-30pt}

\begin{IEEEbiography}[{\includegraphics[width=1in,height=1.25in,clip,keepaspectratio]{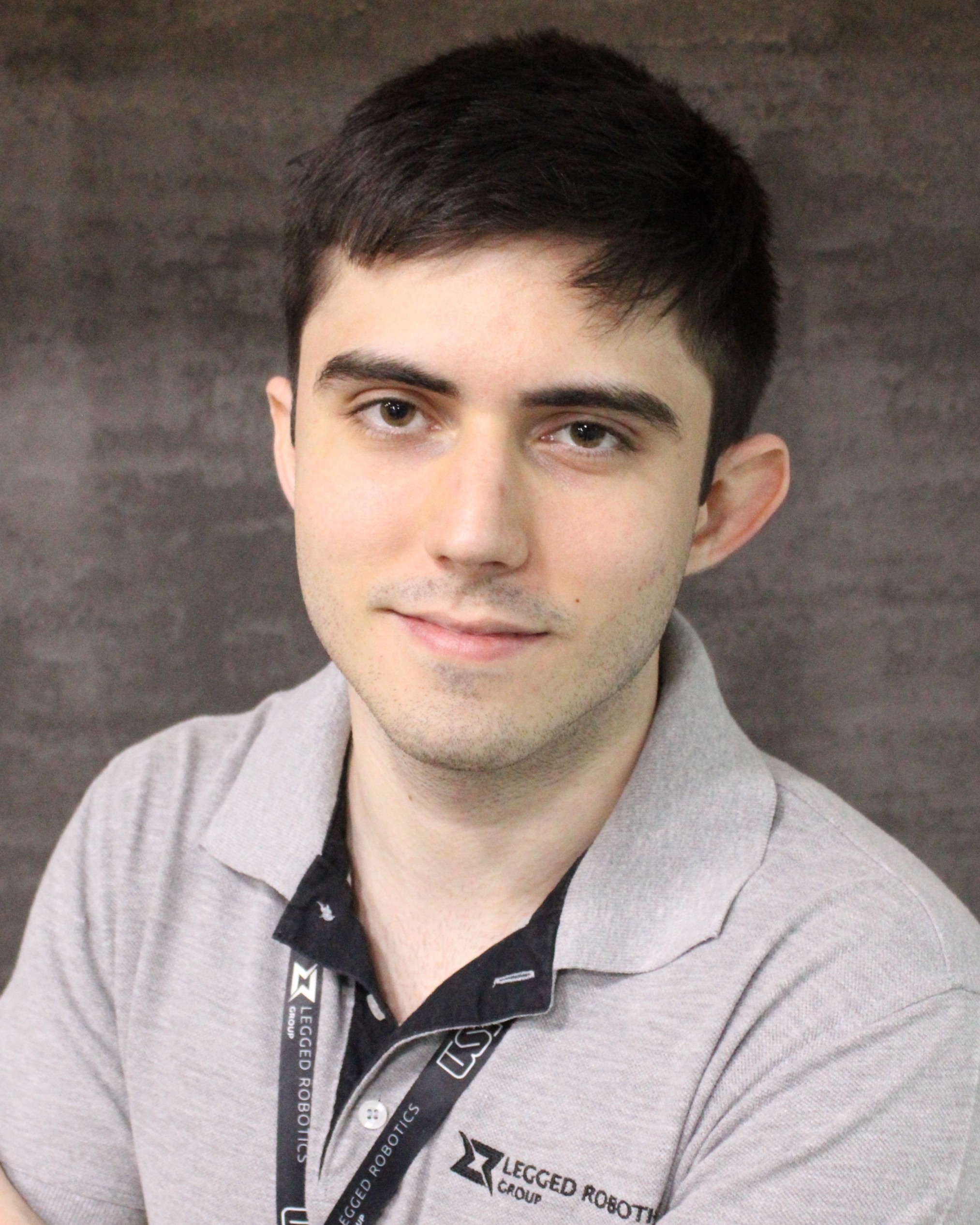}}]{Lucca Maitan}
holds a degree in Mechatronics Engineering from the University of São Paulo (2021). He has experience in the areas of Robotics, Mechatronics, Automation, and software development. During his undergraduate studies, he worked on scientific initiation projects and his final thesis at the Robotic Rehabilitation Laboratory of the Department of Mechanical Engineering at EESC - USP, focusing on exoskeleton control. Since the second semester of 2023, he has been pursuing a Master's degree in the legged robots laboratory at EESC - USP, specializing in control of electric and hydraulic actuators and impedance control for robotic legs.
\end{IEEEbiography}
\vfill

\newpage
\vspace{-33pt}
\begin{IEEEbiography}[{\includegraphics[width=1in,height=1.25in,clip,keepaspectratio]{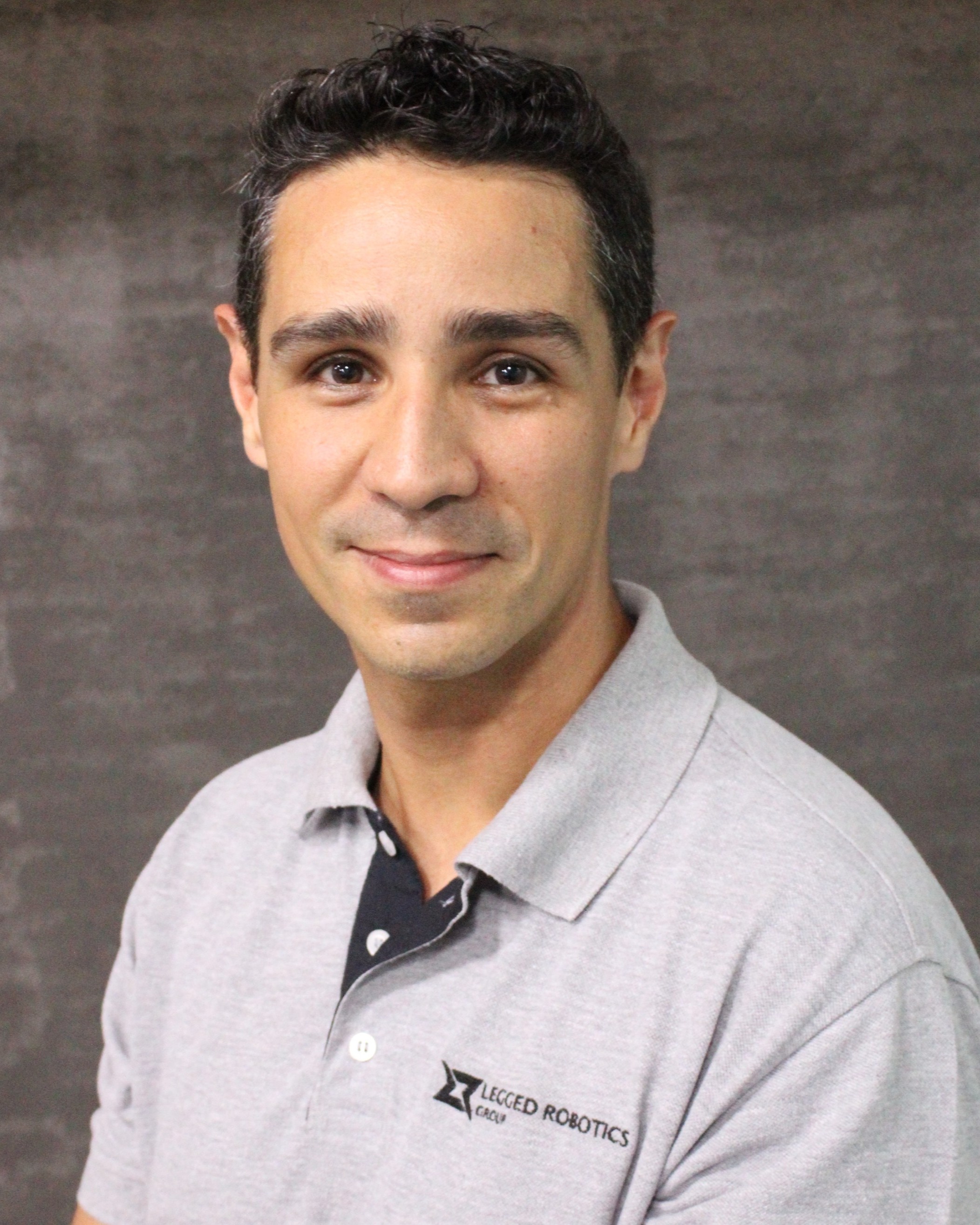}}]{Thiago Boaventura}
received his B.Sc. and M.Sc. degrees in Mechatronics Engineering from the Federal University of Santa Catarina, Brazil, in 2009. He received his Ph.D. degree in Robotics from a partnership between the Italian Institute of Technology and the University of Genoa in Italy in 2013. Then, he joined as a post-doctoral researcher at the Agile Dexterous Robotics Lab at ETH Zurich in Switzerland. There, he was mainly involved in the EU FP7 BALANCE project with a focus on the collaborative impedance control of exoskeleton robots. Thiago is an Assistant Professor at the São Carlos School of Engineering of the University of São Paulo. He is the founder and director of the Legged Robotics Group, a branch of the Robotics Laboratory. He has a strong background in model-based force and impedance control, as well as in hydraulic actuation.
\end{IEEEbiography}
\vfill

\end{document}